\documentclass[sigconf]{acmart}

\usepackage{bm}
\usepackage{algorithm}
\usepackage{algorithmic}
\usepackage{multirow}
\usepackage{ulem}

\AtBeginDocument{%
  \providecommand\BibTeX{{%
    \normalfont B\kern-0.5em{\scshape i\kern-0.25em b}\kern-0.8em\TeX}}}

\setcopyright{acmcopyright}
\copyrightyear{2018}
\acmYear{2018}
\acmDOI{XXXXXXX.XXXXXXX}

\acmConference[WWW'2024]{The Web Conference}{May 13 - May 17, 2023}{Singapore}
\acmPrice{15.00}
\acmISBN{978-1-4503-XXXX-X/18/06}

\begin{document}

\title{RimiRec: Modeling Refined Multi-interest in Hierarchical Structure for Recommendation}

\author{Haolei Pei}
\authornote{Both authors contributed equally to this research.}
\authornote{Corresponding author.}
\affiliation{%
  \institution{NetEase, Inc.}
  \city{}
  \country{}
  }
\email{haleypei@126.com}

\author{Yuanyuan Xu}
\authornotemark[1]
\authornotemark[2]
\authornote{Work done during an internship at NetEase.}
\affiliation{%
  \institution{Southeast University}
  \city{}
  \country{}
  }
\email{yuanyuanxu005@gmail.com}

\author{Yangping Zhu, Yuan Nie}
\affiliation{%
  \institution{NetEase, Inc.}
  \city{}
  \country{}
  }
\email{hzzhuyangping@corp.netease.com}
\email{hznieyuan@corp.netease.com}

\renewcommand{\shortauthors}{Pei and Xu, et al.}

\begin{abstract}
  Industrial recommender systems usually consist of the retrieval stage and the ranking stage, to handle the billion-scale of users and items. The retrieval stage retrieves candidate items relevant to user interests for recommendations and has attracted much attention. 
  Frequently, a user shows refined multi-interests in a hierarchical structure. For example, a user likes \textit{Conan} and \textit{Kuroba Kaito}, which are the roles in hierarchical structure ``\textit{Animation, Japanese Animation, Detective Conan}''. 
  However, most existing methods ignore this hierarchical nature, and simply average the fine-grained interest information.
  Therefore, we propose a novel two-stage approach to explicitly modeling refined multi-interest in a hierarchical structure for recommendation. 
  In the first hierarchical multi-interest mining stage, the hierarchical clustering and transformer-based model adaptively generate circles or sub-circles that users are interested in. 
  In the second stage, the partition of retrieval space allows the EBR models to deal only with items within each circle and accurately capture users’ refined interests. 
  Experimental results show that the proposed approach achieves state-of-the-art performance. Our framework has also been deployed at Lofter.
\end{abstract}

\ccsdesc[500]{Information systems~Recommender systems}

\keywords{Recommender Systems, Multi-interest Recommendation, Generative Retrieval, Embedding-based Retrieval}



\maketitle

\section{Introduction}

\begin{figure}[h]
  \centering
  \includegraphics[width=0.8\linewidth]{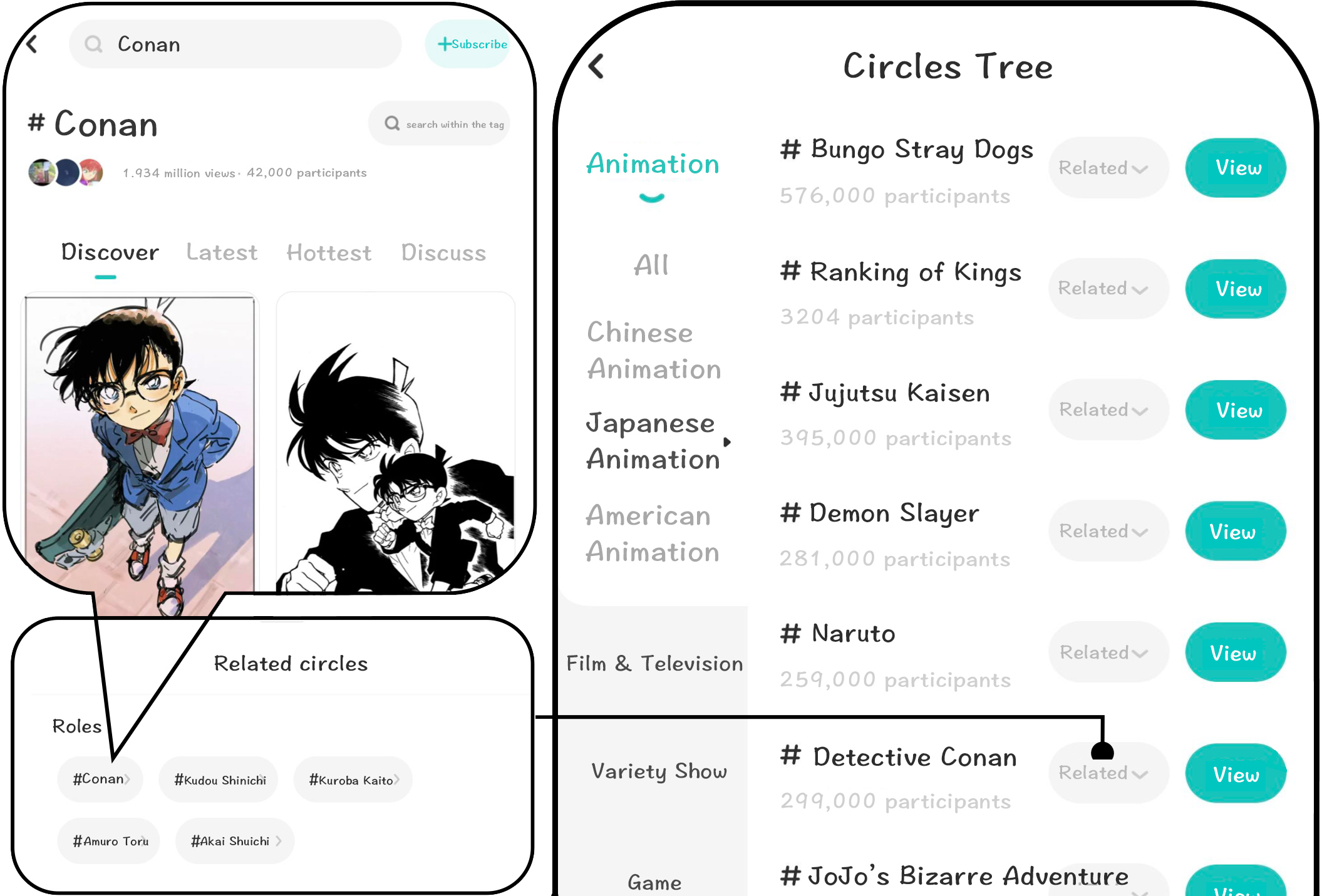}
  \caption{The circle ``\textit{Animation}'' with sub-circles and an content flow example in sub-circle ``\textit{Animation, Japanese Animation, Detective Conan, Conan}''. Note that each circle (including sub-circle) has a content flow and the content of the upper layer circle is a mixture of the contents of sub-circles.}
  \label{Fig.intro_fig}
\end{figure}

Lofter, one of the largest derivative content communities, has over 10 million monthly active users who interact with more than one billion posts. 
A circle, established spontaneously by users, is a community where users share their one specific intimation and interest.
Take Lofter as an example, Figure~\ref{Fig.intro_fig} shows circles  ``\textit{Animation}'', ``\textit{File\&Television}'', ``\textit{Game}'' and so on, for sharing different specific interests. Figure~\ref{Fig.intro_fig} also shows ``\textit{Animation}'' with its sub-circles in different levels, for various interests in different animations of different countries. 
Frequently, a user shows refined multi-interests in a hierarchical structure. For example, a user likes \textit{Conan} and \textit{Kuroba Kaito}, which are the roles in hierarchical structure ``\textit{Animation, Japanese Animation, Detective Conan}''. 
This requires modeling a user's refined interest at a more fine-grained level.

Recently, most prior work for multi-interest recommendation has inferred multiple high dimensional embeddings to represent a user's interests~\cite{CenZZZYT20,PinnerSage,Multi-Interest-Shopping,IHN-RR}, supporting retrieval from the entire library, overcoming the bottleneck of using a single generic user embedding.
Others explicitly partition the library and retrieve items at all sub-libraries with embeddings produced by the same user sequence to extend multi-interest recommendation~\cite{Divide-and-Conquer}. 
However, their approaches suffer from weakness: ignoring this hierarchical nature, and weak to model refined information due to the averaging of more detailed interest information into a limited set of interests.



To address the issues mentioned above, we propose RimiRec, a novel two-stage retrieval method for refined multi-interest recommendation in a hierarchical structure. 
In the first hierarchical multi-interest mining stage, 
a tree for all items was constructed by hierarchical clustering. Each item obtains a Semantic Category ID to model the cycle or sub-circle it belongs to. 
After that, we build a transformer-based model with early stopping to directly generate Semantic Category IDs for interests at different levels based on the user behavior. 
In the second refined multi-interest retrieval stage, the library is explicitly partitioned into sub-libraries based on the item tree. For each user, we capture the items (the user consumed) that fall into the Categories generated in the first stage to build multiple new user sequences, and then train multiple embeddings to perform ANN searches in sub-libraries of correspondent Categories respectively. This explicitly partitioned method allows longer behavior sequences in each category for modeling a user's personalized needs in refined interests on various levels.

Our main contributions are as follows:
\begin{itemize}
    \item A novel two-stage retrieval model for refined multi-interest recommendation in a hierarchical structure was proposed, contributing to mine users’ interests on various levels.
    \item The transformer-based model generates circles on different levels flexibly, which is the key to hierarchical multi-interest.
    \item Our approach achieves state-of-the-art performance, excelling in both offline datasets and real-world online A/B testing scenarios. Provide industrial solutions for hierarchical multi-interest recommendations in similar scenarios.
\end{itemize}

\section{Method}

\subsection{Overall Architecture} 
Suppose there are a set of users $\mathcal{U}$ and a set of items $\mathcal{I}$. 
Denote the user-side features as $\bm{x}_u$ for each user $u \in \mathcal{U}$ and item-side features as $\bm{x}_i$ for each item $i \in \mathcal{I}$. 
Embedding-based retrieval (EBR) methods use a user encoder~$f$ and an item encoder~$g$ to transform user and item features into embeddings~$\bm{e}_u = f(\bm{x}_u)$ and $\bm{e}_i = g(\bm{x}_i)$ respectively, where $i \in [1:\lvert \mathcal{I} \rvert]$. 
The distance (most commonly inner product) between user embedding $\bm{e}_u$ and item embedding $\bm{e}_i$ is used to capture the relevance of item $i$ to user $u$ denoted by $r_{ui}$.

\begin{figure}[h]
\centering
\includegraphics[width=\linewidth]{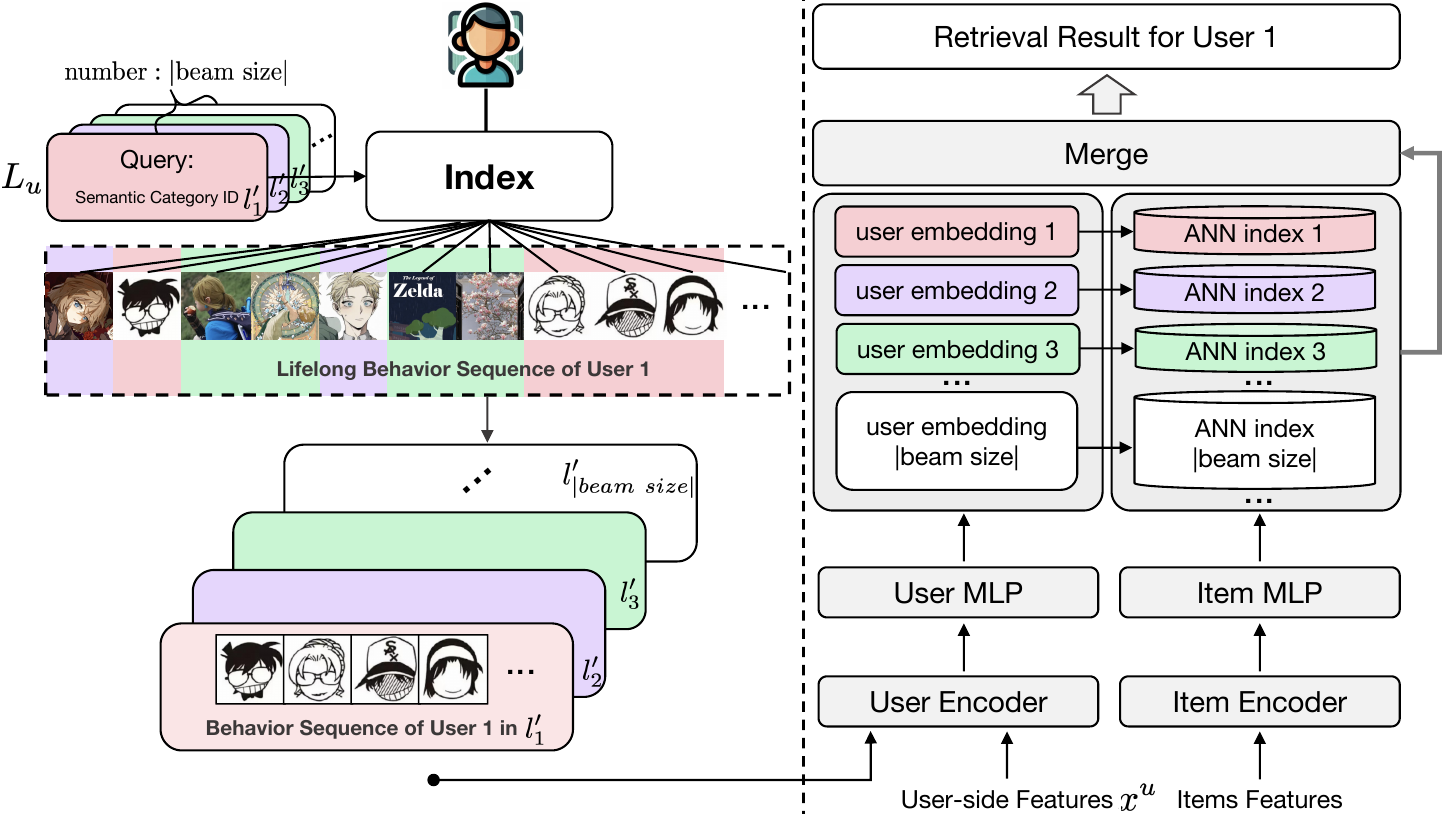}
\caption{The architecture of the item retrieval stage of the proposed approach, RimiRec.}
\label{Fig.overview-stage2}
\end{figure}

As shown in Figure~\ref{Fig.overview-stage1} and Figure~\ref{Fig.overview-stage2}, our proposed method consists of two components: 
(1) The hierarchical multi-interest mining stage generated interests ID at different levels for users by a generative model; 
(2) The refined multi-interest retrieval stage performs ANN searches in partitioned retrieval space with corresponding embeddings to achieve refined and personalized recommendations.

\subsection{Hierarchical Multi-interest Mining}
\label{sec_Multi-interest Mining}
\textbf{Semantic Category ID Assignment. }
Semantic Category ID was assigned for each item based on the item features (e.g. titles, descriptions), representing the circle the item belongs to.
Specifically, map the item features to embedding vectors and then cluster via the hierarchical $k$-means algorithm. As shown in Algorithm~\ref{alg:k-means}, all items are first clustered into $k$ clusters. For clusters with more than $c$ items, the $k$-means algorithm is applied recursively. 
Each cluster in the $i$-th layer is assigned a number $r_i$ starting from 0 to at most $k$-1. 
In this way, we organize all items into a tree structure, and the semantic category ID set $L$ of the leaf circles is obtained, where the category of each leaf circle in $L$ is represented by path $l=\{r_1,\dots,r_m\}$. Futhermore, each item in the leaf cluster $r_m$ in $l$ is assigned any path $l'=\{r_1,\dots,r_p\}$, where $1 \leq p \leq m, p \in \mathbb{N}^+$, as the semantic category ID.  

\begin{figure}[h]
\centering
\includegraphics[width=\linewidth]{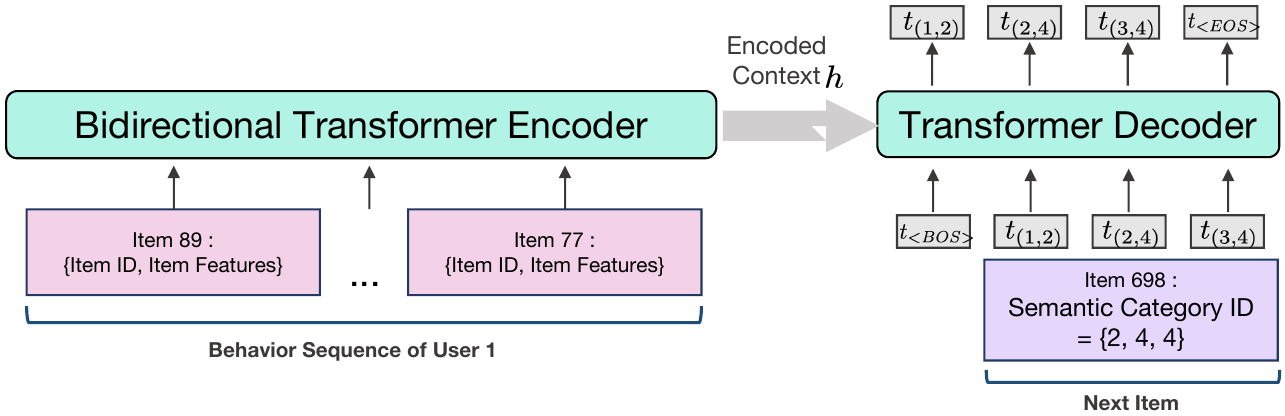}
\caption{The architecture of the multi-interest mining stage of the proposed approach, RimiRec.}
\label{Fig.overview-stage1}
\end{figure}

\renewcommand{\algorithmicrequire}{\textbf{Input:}}
\renewcommand{\algorithmicensure}{\textbf{Output:}}
\begin{algorithm}[h!] 
\begin{algorithmic}[] 
\REQUIRE ~~ \\
    Item embedding $\bm{e}_{1:\lvert \mathcal{I} \rvert}$\\
    Number of clusters $k$\\
    Recursion terminal condition  $c$ \\
    Prefix $l_{prefix}$ (Initialized to an empty string)\\
\ENSURE ~~\\ 
 Semantic category ID $L$ of leaf circles\\
\renewcommand{\algorithmicensure}{\textbf{Function: GenerateSemanticCategoryID}($\bm{e}_{1:\lvert \mathcal{I} \rvert}$, $l_{prefix}$)}
\ENSURE ~~\\ 
\STATE $C_{1:k}$ $\leftarrow$ $KMeansCluster(\bm{e}_{1:\lvert \mathcal{I} \rvert}, k)$
\STATE $L$ $\leftarrow$ $\varnothing$
\FOR{i $\in$ [0, $k-1$]}
    \STATE $r_i$ $\leftarrow$ $i$ \\
    \IF{$|C_{i + 1}|$ > $c$}
        \STATE $l$ $\leftarrow$ \textbf{GenerateSemanticCategoryID}$(C_{i + 1}, l_{prefix} + str(i))$ \\
   \ELSE
        \STATE $l$ $\leftarrow$ ConcatString($l_{prefix}, r_i$) \\
    \ENDIF
    \STATE $L$ $\leftarrow$ $L$.Append($l$)\\
\ENDFOR \\
\RETURN $L$
\end{algorithmic}
\caption{Hierarchical $k$-means.}
\label{alg:k-means}
\end{algorithm}

\textbf{Multi-interest Generation. }
A Transformer-based sequence-to-sequence model is trained on the user behavior sequences $\bm{s}^u$ to predict the possible circles that the user is interested in by generating corresponding semantic category IDs. Note that $\bm{s}^u$, including the sampled long-term sequence and all short-term sequence items, is obtained from the user's lifelong behaviors $\mathcal{I}^{u}$.
The generative-based model can adaptively learn the users' refined interest reflected in generating IDs of different lengths via an early stopping mechanism (that generates the end token).
Notice that we did not share the embedding space of the encoder and decoder since the vocabulary spaces of the user’s interaction history and semantic category IDs are quite different. 
The implementation details are as follows.

Given a user behavior sequence $\bm{s}^u$, the probability of generating a semantic category ID $l'$ is calculated as follows: 

\begin{equation}
    h = \text{TransformerEncoder}(\bm{s}^u;\theta_{encoder}) 
\label{h_1}
\end{equation}
\begin{equation}
    P(l'|\bm{s}_u,\theta) = \prod_{i=1}^{p}{P(r_{i}|h, r_{1}, r_{2}, ..., r_{i-1}, \theta_{decoder})}
\label{p_1}
\end{equation}

where $h$ is the representation output from encoder; $\theta$ denotes the whole model parameters.
In the training stage, $p = m$ represents the semantic category ID of the leaf circle to which the items that the user $u$ has interacted historically belong. In the inference stage, we perform a beam search on the decoder network with a prefix tree for valid identifiers, to obtain the user's refined interest $L_{u} = \{l'_1,\dots,l'_{\lvert \text{beam size} \rvert}\}$, where $1 \leq p \leq m, p \in \mathbb{N}^+$.

In particular, the same number appearing at different positions in a semantic category ID is different.
For instance, as shown in Figure~\ref{Fig.overview-stage1}, the token ``$t_{(2,4)}$'' and ``$t_{(3,4)}$'' are different tokens in the vocabulary space, corresponding to the circle IDs in layer2 and layer3, respectively. Together (with token ``$t_{(1,2)}$'') they represent the the semantic category ID ``$\{2, 4, 4\}$''.

\subsection{Refined Multi-interest Retrieval}
A modified two-tower model was used for multi-interest retrieval.
First partition the whole library $\mathcal{I} $ into sub-libraries $\{\mathcal{I}_1,\dots,\mathcal{I}_{\lvert L \rvert}\}$ based on the semantic category ID set of the leaf circles $L$ defined in Section~\ref{sec_Multi-interest Mining}. Note that sub-library $\mathcal{I}_i$ can be merged at the upper layer flexibly based on $L_{u}$ generated in the first stage, and $\{\mathcal{I}^1,\dots,\mathcal{I}^{\lvert L_{u} \rvert}\}$ are obtained. 
Secondly, the user's lifelong behavior sequences $\mathcal{I}^{u}$ is split following the Key-Key-Value data structure: the first key is the user ID, the second keys are semantic category IDs $L_{u}$ generated in the first stage and the last values are the user behavior items $\mathcal{I}^u_{l'_{n}}$ that fall into each semantic category ID $l'_{n}$. This allows a longer behavior sequence for each category compared to the previous method, and massive noise existing in the lifelong user behavior that hinders refined interest modeling can be filtered by the split strategy.
After that, for each Key-Key-Value, with a separate sequence, a more refined user embedding is learned, and further a user has multiple distinction user embeddings. Then retrieve relevant items in the sub-library $\mathcal{I}^i$ that each user embedding corresponds to instead of the whole library. 
Lastly, the beam probability in the first stage will be used for assigning quotas to merging results.
We only use items from the same category of positive items as negative items (that are mostly hard negatives) for training as well. 
The training loss is 
\begin{equation}
\label{loss_intag_neg}
    \mathcal{L} =  - \sum_{u \in \mathcal{U}} \sum_{n=1}^{\lvert L_u \rvert} \sum_{i^{+} \in \mathcal{I}^u \cap \mathcal{I}^{n}} \Bigl[\log(\sigma(r_{ui^+})) + \mathop{\mathbb{E}}_{i^-\sim \mathcal{I}^{n} \setminus \mathcal{I}^u}\bigl[\log(1-\sigma(r_{ui^-}))\bigr]\Bigr].
\end{equation}

where $\mathcal{I}^{n}$ represents the sub-library of the semantic category ID $l'_{n}$.
During employment, item embeddings $\{\bm{e}_i\}_{i\in \mathcal{I}}$ are calculated beforehand in an offline or near-line system and indexed by approximate nearest neighbor (ANN) search system (e.g. FAISS~\cite{johnson2019billion}). Therefore, only the user encoder $f(\mathcal{I}^u_{l'_{n}};\bm{x}_u)$ needs to be evaluated in real-time, and the ANN system can efficiently retrieve the nearest items in sub-linear time. 

The partition of item space allows deal only with items within each circle and accurately captures users’ refined interests in the target circle, meeting users’ different needs in different circles.

\section{Offline Experiments}

\textbf{Datasets.}
We conduct extensive experiments on two public datasets. Table~\ref{dataset_statistics} summarizes the detailed statistics of the datasets. 
\begin{itemize}
    \item \textbf{MovieLens-1M}~\cite{HarperK16} is a public movie rating dataset commonly used in recommender systems. Randomly sample 80\% of the users as the training set and 20\% of the users as the test set. Learn only from the user sequences without user ID.
    \item \textbf{LofterData} collects user behaviors from the logs of Lofter for 15 days with 12,000 users (that clicks and daily interactions are less than $20$ and $5$ respectively were filtered out). The 15th-day data is retained for testing.
\end{itemize}

\textbf{Parameter settings.}
We set $k$ as 4 and 6, c as 250 and 100,000 for ML-1M and Lofterdata respectively.
During inference, each method generates a candidate set of $M$ relevant items and we evaluate its Recall@$M$ and HitRate@$M$ on the following $N$ disjoint behaviors. 
We set $M \in \{20, 50\}$, $N=10$ for ML-1M and $M \in \{500, 1000\}$, $N=25.3$ in average for LofterData. 

\textbf{Baselines.} We compare our approach with the common baselines used in the industry:
\textbf{MIND}~\cite{LiLWXZHKCLL19} uses capsule networks with the routing mechanism to group user behaviors into multiple clusters and obtain multiple embeddings for retrieval. \textbf{ComiRec-SA}~\cite{CenZZZYT20} uses multi-head attention mechanisms to generate multiple embeddings to capture their diverse interests. \textbf{Divide\&Conquer}~\cite{Divide-and-Conquer} divides the whole set of items into multiple sub-libraries and runs EBR to retrieve relevant candidates from each sub-library in parallel.

\begin{table*}[t]
\caption{Performance comparison on MovieLens-1M and LofterData. All the numbers are percentage numbers with ‘\%’ omitted.}
\label{comparsion}
\renewcommand\arraystretch{0.9}
\begin{tabular}{@{}ccccccccc@{}}
\toprule
\multirow{2}{*}{Method} & \multicolumn{4}{c}{MovieLens-1M} & \multicolumn{4}{c}{LofterData} \\ \cmidrule(lr){2-5} \cmidrule(lr){6-9}
 & Recall@20 & HitRate@20 & Recall@50 & HitRate@50 & Recall@500 & HitRate@500 & Recall@1000 & HitRate@1000 \\ \midrule
MIND & 4.557 & 23.372 & 11.605 & 43.294 & 24.754 & 61.158 & 38.719 & 71.660 \\
ComiRec & 7.744 & 44.238 & 17.165 & 65.820 & 26.010 & 61.467 & 40.410 & 70.156 \\ 
Divide\&Conquer & 7.107 & 48.014 & 15.465 & 69.889 & 27.722 & \textbf{68.035} & 44.965 & \textbf{75.95} \\ 
RimiRec-2beams & \textbf{9.119} & \textbf{50.878} & \textbf{19.263} & \textbf{73.542} & \textbf{32.428} & 66.332 & \textbf{46.581} & 72.781  \\ 
RimiRec-4beams & 9.401 & 51.975 & 20.498 & 76.176 & 37.083 & 66.561 & 59.376 & 73.667 \\ 
RimiRec-6beams & 9.461 & 52.571 & 20.630 & 76.207 & 39.978 & 67.740 & 65.430 & 73.748 \\ 
\midrule
Improvement & +17.8\% & +15.0\% & +12.2\% & +11.7\% & +24.7\% & +7.9\% & +15.3\% & +3.7\%\\ \bottomrule
\end{tabular}
\end{table*}

\textbf{Main Result.}
Table~\ref{comparsion} summarizes the performance of the proposed approach as well as baselines on MovieLens-1M and LofterData. First, the proposed approach outperforms all the baseline methods on Recall consistently on both datasets and achieves up to $17.8\%$ and $24.7\%$ improvements over the baseline model ComiRec on MovieLens-1M and LofterData datasets, respectively. However, our HitRate is not as good as Divide\&Conquer. The reason could be that Divide\&Conquer searched in almost all sub-libraries. However, our method achieves better performance under the trade-off in Recall and HitRate.
Secondly, as the number of interest embeddings increases, our model performance becomes better. 
An \textbf{Ablation Study} was done to assess the proposed RimiRec approach. Results in table~\ref{ablation_study} show that by performing ANN search in the whole library rather than in the corresponding sub-libraries, the Recall@500 and Recall@1000 performance drops by up to $-8.0\%$ and $-4.5\%$ respectively on LofterData. What's more, results show that performing ANN search with user embedding produced by a unified user sequence rather than the sequence under the corresponding category, the overall performance drops by up to $-4.4\%$ and $-2.3\%$ on LofterData, respectively. The reason why the improvement is not significant could be that the sequence length in specific categories on the offline data set is shorter and the features are sparser.

\begin{table}[t]
\caption{Statistics of datasets (after preprocessing).}
\label{dataset_statistics}
\renewcommand\arraystretch{0.9}
\begin{tabular}{@{}cccccc@{}}
\toprule
Dataset & \#Users & \#Items & \#Interactions &  \#Leaf Circles \\ \midrule
ML-1M & 6,040 & 3,411 & 892,458 & 28 \\
LofterData & 12,000 & 798,949 & 808,619 & 176 \\ \bottomrule
\end{tabular}
\vspace{-0.4cm}
\end{table}

\begin{table}[t]
\caption{Ablation study on the Lofterdata (with ‘\%’ omitted).}
\label{ablation_study}
\renewcommand\arraystretch{0.9}
\begin{tabular}{@{}ccccc@{}}
\toprule
Method & Recall@500 &Recall@1000\\ \midrule
\multicolumn{1}{l}{w/o sub-libraries (2beams)} & 30.028 & 44.590 \\
\multicolumn{1}{l}{w/o behavior splitting (2beams)} & 31.061 & 45.534 \\ \midrule
\end{tabular}
\vspace{-0.4cm}
\end{table}

\section{Online A/B Experiments}
\begin{figure}[h]
\centering
\includegraphics[width=\linewidth]{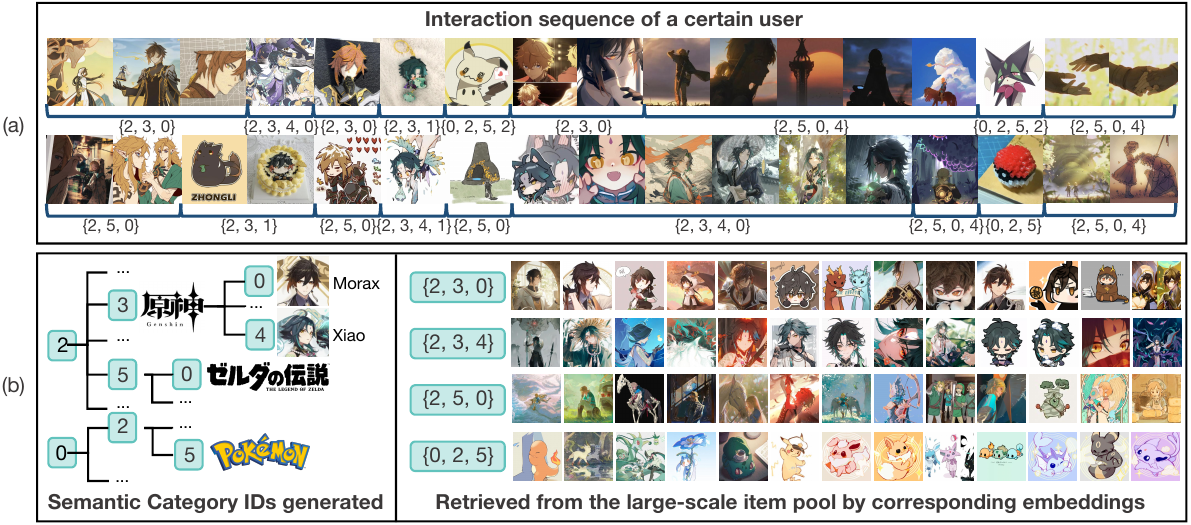}
\caption{(a) Part of the behavior of a certain user in Lofter; (b) Left: The Semantic Category IDs of interests in role-levels and IP-levels were generated. Right: the items retrieved by the Refined Multi-interest Retrieval stage. Specific roles and other IP-related content were recommended personalized.}
\label{Fig.case_study}
\end{figure}

The proposed approach has also been tested in a strict online A/B experiment at Lofter for a month (from 2023-08-03 to 2023-09-03), and involved over 10\% users in the experiment group, yielding statistically significant at $p < 0.05$ for all experimental results.
In the experiment group, the proposed retrieval method serves as one of the candidate sources in the candidate generation phase. Compared with the control group (that ComiRec was deployed), our method leads to significant improvements on most of the key online metrics as shown in Table~\ref{ab-test}. 
Currently, RimiRec has been deployed online and serves the main scene daily.
Figure~\ref{Fig.case_study} shows an example result of RemiRec. The model captures the user’s refined multi-interests in a hierarchical structure. Especially clearly captures role-level interest that previous methods fail to capture.

\begin{table}[t]
\caption{Results of the online A/B experiment. Note that: 1) Diversity indicates the Ratio-based diversity~\cite{Surrogate-User-Experience} with its numerator replaced by the number of unique items the user consumed, measuring the diversity and broadness of the contents that a user consumed. 2) User Activity records the frequency of users visiting the app in a week.}
\label{ab-test}
\renewcommand\arraystretch{0.9}
\begin{tabular}{@{}ccccc@{}}
\toprule
User Activity & Click & Engagement & Retention & Diversity \\ \midrule 
+0.31\% & +0.76\% & +2.00\% & +0.45\% & +1.26\% \\ \bottomrule
\end{tabular}
\vspace{-0.1cm}
\end{table}

\section{Conclusion}
This paper proposes a production-proven two-stage retrieval model for refined multi-interest recommendation in a hierarchical structure. 
The hierarchical clustering and transformer-based model adaptively generate circles or sub-circles that users are interested in. 
The partition of retrieval space allows the EBR models to only deal with items within each circle and accurately capture users’ refined needs in different interests on various levels. 
Extensive results in offline and online A/B experiments demonstrate the effectiveness of the proposed solution.

\bibliographystyle{ACM-Reference-Format}
\bibliography{sample-base}

\end{document}